\documentclass[twocolumn]{article}
\usepackage{epsfig}

\title{Stochastic resonance and the trade arrival rate of stocks}
\author{A. Christian Silva \\
Department of Physics\\
 University of Maryland, College Park, Maryland 20742\\
a.christian.silva@gmail.com\\
 and\\
 Ju-Yi J. Yen\\
Department of Mathematics\\
Vanderbilt University, Nashville, Tennessee 37240\\
 ju-yi.yen@vanderbilt.edu}

\begin{document}
\maketitle


\begin{abstract}
We studied non-dynamical stochastic resonance for the number of
trades in the stock market. The trade arrival rate presents a
deterministic pattern that can be modeled by a cosine function
perturbed by noise. Due to the nonlinear relationship between the
rate and the observed number of trades, the noise can either enhance
or suppress the detection of the deterministic pattern. By finding
the parameters of our model with intra-day data, we describe the
trading environment and illustrate the presence of SR in the trade
arrival rate of stocks in the U.S. market.
\end{abstract}

\section{Introduction}
It can be very useful to model and understand market microstructure. For instance,  it has been shown that the number of trades and volume can help
explain volatility fluctuations and therefore help model asset returns \cite{Ane2000,Clark1973,Silva2007}. More importantly, these studies link market activity (and therefore market microstructure) to
the large class of asset allocation and pricing models based on time changed Brownian motion \cite{Yor2002,Yen2008,Schoutens2003}.  More practically, we also point to the increase of intra-day trading that some claim
should account for more than $50\%$ of the trading volume by 2010. High-frequency trading will increasingly  require market microstructure for risk control and trading.

In this paper, we use the doubly stochastic Poisson process to model the number of trades inside of a trading day. Using this model, we focus on the interaction between noise and the trade rate "U"-shaped pattern (higher average arrival rate at the open and close of the trading day). We report an interesting feature of such model: stochastic resonance (SR). The presence of stochastic resonance indicates that a correct amount of noise is, in fact, beneficial to increase the detection of the "U"-shaped pattern. That is, random trading (represented by random arrival rates) can constructively contribute to the existence of such periodic patterns.
\subsection{"U"-shaped pattern}
Finance and economics show a variety of well documented periodic patterns \cite{Franses1996}. In particular, there is the intra-day "U"-shaped pattern (or patterns).
\footnote{Also sometimes called the intra-day "smile" or inverse J pattern.} That is, the average intra-day volatility,
the trading volume, bid-ask spreads, trading activity (number of trades), and even returns are significantly higher near the open and close of the trading
day \cite{Walsh1999,Lee2001,Linn1990,Osborne1962,Ord1985}. These quantities form on average an "U" shape; hence, the name "'U'-shaped pattern".
In fact, it has been shown that there need not be a official exchange halt.
Studies using 24-hour markets (such as currency markets) find that "U"-shaped patterns are also present within the normal business hours \cite{Winters2004}.

There is a large literature that explains the different "U"-shaped patterns. In general the literature has been grouped into two lines:
The first line suggests that asymmetric information is a necessary condition for the "U" shape to exist. By asymmetric
information, it is meant that some traders have private information about the future asset returns \cite{Pfleiderer1989} or
payoffs \cite{Wang2000}. It is the interplay between informed traders and uniformed traders together with market stops
that will lead to the observed pattern both in returns and volumes \cite{Wang2000}.

The second line contests the importance of asymmetric information by pointing out to the presence of the same patterns in markets with no
asymmetric information \cite{Winters2004,Winters2001}. They argue that the presence of market stops could be enough. This line emphasizes
the exchange of risk close to market stops \cite{Mulherin1992}. The argument is that information accumulated overnight will generate an
increase of the trading activity at market open because portfolio managers will want to bring their portfolio to a particular risk target as soon
as possible. The same will happen at the end of a trading day, when traders will want to prepare themselves for the next overnight period.
Therefore, the pattern could be just due to the timing of transaction decisions by institutional investors and long-term investors with short term
investors representing the other side of the trade \cite{Maberly2000}.

In this paper we do not address the micro-economical origin of such patterns as, for instance, in Hong and Wang \cite{Wang2000}. We take an empirical approach to the problem. We start with a plausible statistical model which describes the data well. Since our model presents stochastic resonance, we conjecture that such a phenomenon can be important to the "U" pattern. The presence of SR could indicate that noise traders can, in fact, magnify such pattern. However, in the present paper we will not be able to reach such level of description, leaving such questions for future studies.

From a practical point of view, our work still presents useful results. For instance, one can also think on a possible algorithm where random trading (hence undetected to other traders) can enhance even further such a pattern. Of course, the benefit of enhancing such a pattern is, a priori, not clear. But in a similar fashion there could be other patterns where similar analysis will lead to undetected (and beneficial to the trader) market manipulation. We also point to the analogy drawn recently by Dempster et al. \cite{Dempster2008} as a possible starting point to generalize SR to asset allocation, for instance.
\subsection{Stochastic resonance}
Stochastic resonance (SR) has been successfully used to explain signal transmission and detection in a
wide variety of systems in diverse areas such as physics, geology,
neural science, engineering, and biology \cite{Marchesoni1998}. SR is briefly defined as noise-enhanced signal detection.
Typically there is a threshold below which the signal cannot be detected. Adding noise to the signal can increase
the number of events that cross such a threshold, hence increasing the detection of such signal. Imagine a sine wave of amplitude less than the detection barrier.
Noise added to the sine wave will increase the events that cross the detection barrier. However, if the noise intensity is too high, it will corrupt the signal beyond recognition.

In finance, stochastic resonance has been suggested to explain market crash and high-frequency patterns in the exchange rate \cite{Holyst2003,Sato2006}.
Different from previous work, our paper employs non-dynamical, threshold-free stochastic resonance \cite{Bezrukov1997,Bezrukov1998}. Non-dynamical SR is purely stochastic
in nature and, therefore, it does not require a microscopic model (such as Newton's equations of motion). Consequently, it can be more general because it dissociates a specific
microscopic model from the SR phenomenon.  To the best of our knowledge, this is the first time this particular SR class has been applied to financial data.

\section{Non-dynamical stochastic resonance}
In direct analogy to non-dynamical SR in other disciplines, we assume that trade
arrivals follow a time-dependent Poisson process. We take the point
of view of a market participant and observer. We observe signals in
the market that are being transmitted to all market participants.
Our goal is to describe the present state of the trading
environment and therefore detect the presence of SR.

Non-dynamical and threshold-free stochastic resonance was originally
discovered for systems that can be modeled with a doubly stochastic
Poisson process (or time-dependent Poisson process) \cite{Bezrukov1998}. Such a process
is constructed out of Poisson process $N(t)$ with the probability
distribution of $n$ events during time interval $T$ given by

\begin{equation}\label{poisson}
p_N(n)=\frac{e^{-r T} (r T)^n}{n!},
\end{equation}

\noindent where the event arrival rate $r$ is itself a time
dependent random process independent of $N(t)$ \cite{Korolev2002}. Bezrukov
and Vodyanoy have shown that the following arrival rate $r(t)$ of a
doubly stochastic Poisson process exhibits SR:

\begin{equation}\label{Rt}
r(t)=r(0)\exp (A(t)),
\end{equation}

\noindent with $A(t)$ as

\begin{equation}\label{Vt}
A(t)=W(t)+A_s\cos(2\pi f_st).
\end{equation}
That is, $A(t)$ is the sum of zero mean Gaussian noise $W(t)$ with
root mean square $\sigma$; and a weak periodic signal with amplitude
$A_s$ and frequency $f_s$ \cite{Bezrukov1998}.

In the case of Equation~(\ref{Vt}), the slow cosine-wave
$A_s\cos(2\pi f_st)$ can be better detected, due to SR, by analyzing
the final Poisson generated time series $N(t)$ if there is a
correctly tuned noise $W(t)$. However, excessive noise might corrupt
the signal.

\subsection{Signal-to-noise ratio}
For a quantitative analysis of the synchronized process,
signal-to-noise ratio (SNR) is introduced to measure stochastic
resonance. SNR is defined as the height of the signal divided by the
height of the noise background of the power spectrum at the signal
frequency $f_s$ \cite{Marchesoni1998}. We speak of SR if there is a maximum in
the output SNR as a function of noise intensity $\sigma$. That is,
there is an optimal value of the noise that concurs with the
periodic signal.

Bezrukov and Vodyanoy derived the output SNR for the doubly
stochastic Poisson process with the rate given by
Equation~(\ref{Rt}) and Equation~(\ref{Vt}) via power spectral
density for $A_s \ll \sigma$ and $f_s$ much smaller than all other
characteristic frequencies.\footnotemark\footnotetext{They have, in
fact, derived it for a sine-wave signal; howeve, the power spectral
densities for sine and cosine are identical. Using cosine instead of
sine here will not change the SNR.} The output SNR is given by

\begin{equation}\label{s2n}
SNR=\frac{(A_{s}^{2}r(0)/2)\exp(\sigma^2 /2)}{2+(2r(0)/\pi
f_c)\exp(\sigma^2/2)\sum_{n=1}^{\infty}(1/n)(\sigma^{2n}/n!)},
\end{equation}

\noindent where $f_c$ is the noise corner frequency \cite{Bezrukov1997,Bezrukov1998}. We
also have $f_c=(2\pi \tau_c)^{-1}$, where $\tau_c$ is the noise
$W(t)$ correlation time. The noise has a Lorentzian power spectrum
with exponential autocorrelation function. We have also verified
Equation~(\ref{s2n}) by Monte Carlo (MC) simulating
Equation~(\ref{poisson}) with rate given by Equations~(\ref{Rt}) and
(\ref{Vt}) \cite{Weron2004}. Using our MC simulations, we remark that
Equation~(\ref{s2n}) is still a very good approximation outside the
range of parameters for which it has been derived \cite{Bezrukov1998}.

The maximum SNR as a function of $\sigma$ in Equation~(\ref{s2n}) is
the result of a competition between the exponential in the numerator
and the exponential times a linear term (to first order
approximation) in the denominator. The constant $A_{s}^{2}r(0)/2$ is
a general scale of the amplitude of the maximum. The location of the
maximum is given by $r(0)/f_c$. $r(0)/f_c$ is the number of events
that occur over the correlation time. If $r(0)/f_c$ is large, the
maximum moves to small $\sigma$; if $r(0)/f_c$ is small the maximum
moves to large $\sigma$. Therefore, the existence of the maximum
depends entirely on $r(0)/f_c$: SR is more important for a random
variable where the statistics is poor; that is, the equilibrium
arrival rate $r(0)$ is small.
SR is also more important when the noise present (or added) to the
system is uncorrelated. Furthermore, it also shows that no matter
how good $r(0)$ is, the output signal detection could be further
improved if $f_c$ is big enough. Equation~(\ref{s2n}) shows that if
the condition $f_c>2r(0)\pi$ [$\tau_c r(0) < 0.25$] holds, the
process demonstrates SR.

\section{Numerical experiments}
We cannot always perform controlled experiments in economics;
therefore, we need to extract all the needed information from the
data we observe. We first parameterize the time dependent rate
$r(t)$ defined by Equations~(\ref{Rt}) and (\ref{Vt}) without the
Gaussian noise $W(t)$ by assuming that parameters for the rate are
constant for at least $1$ month. We have
\begin{equation} \label{rateT}
r(t) = r(0) e^{A_s \cos(2\pi f_s t)},
\end{equation}
\noindent where $f_s$ is one over the trading day, $f_s = 1/23400$
seconds. Notice that once we have detected the periodic signal $A_s
\cos(2\pi f_s t)$, we have bypassed one of the potential benefits of
SR: increasing statistics to help the signal detection. Therefore,
we cannot infer if our signal reconstruction was helped by SR.
However, we can still say that a system can potentially present SR
by identifying all the parameters.

In our numerical experiments, we obtained tick-by-tick financial
data (smallest time resolution is $1$ second) for $12$ trading
stocks (AA,AEP, AES, AMG, AOL, ATI, EK, F, JPM, PFE, PG, and S)
during $21$ trading days (January 2001). First, we analyze the
daily pattern by constructing a proxy for the trade arrival rate
inside of the $6.5$ hours composing the trading day. We bin the data
in $2$-minute ($120$-second) intervals using $1$-second resolution
from the first day to the last day (total of $21$ days). We assume
that the rate in each of the $2$-minute bin is constant. We remove
the first $5$ minutes at the beginning of the trading day due to
data inaccuracy. We end up with a total of 192 $2$-minute intervals
in a day.

We then calculate the arrival rate for every $2$ minutes by counting
the number of trades in every interval and dividing by $120$
seconds; that is, the average number of trades in each of the $192$
bins during $21$ days. This is equivalent to fitting a constant rate
Poisson process since the maximum likelihood estimate for the rate
is simply the average number of events per unit of time. Once we
have the empirical daily rate pattern, we fit the daily rate pattern
to Equation~(\ref{rateT}) to estimate $r(0)$ and $A_s$.

\begin{figure}
\centerline{\epsfig{file=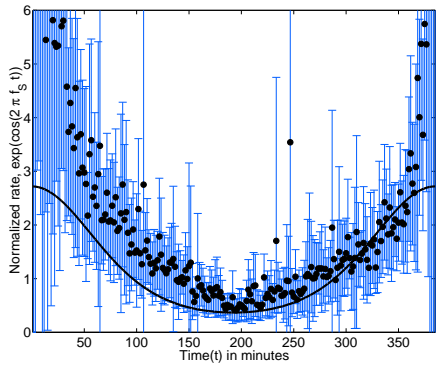,width=0.6\linewidth,angle=-90}}
\caption{\footnotesize\sf Normalized average intra-day arrival rate
pattern for 12 stocks. The number of trades per unit of time
(2-minute bins are used here) are chosen as a good proxy for the
arrival rate of trades. The symbols are the average over 21 days and
12 stocks. The error bars represent 1 standard deviation of the
empirical data. The solid line is the theoretical model in
Equation~(\ref{rateT}).} \label{fig:intraday}
\end{figure}

FIG. \ref{fig:intraday} shows the arrival rate averaged over $21$
days and all $12$ stocks. In order to average over all stocks, we
subtract $\ln(r(0))$ from the natural logarithm of the empirical
rate of each stock and then multiple by $1/A_s$ to obtain $\cos(2\pi
f_s t)$. Thus, we are able to recover the parameter independent
$\exp(\cos(2\pi f_s t))$. Equation~(\ref{rateT}) is a very simple
periodic function and hence the most simple example one can choose.
However, FIG. \ref{fig:intraday} shows that it gives a good
approximation to the empirical daily rate pattern especially for the
second half of the day. The first half is on average systematically
higher indicating that a skewed function could be better.
Nonetheless, since $r(0)$ measures the vertical shift of the whole
curve, the value of $r(0)$ will not change very much even if
asymmetry is taken into account. The quality of our approximation
will suffice for defining SR.

\begin{figure}
\centerline{\epsfig{file=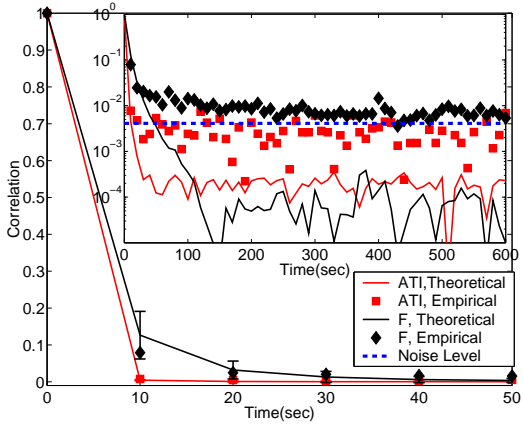,width=0.7\linewidth,angle=-90}}
\caption{\footnotesize\sf Autocorrelation of the number of trades
for $2$ stocks (ATI and F). The symbols represent the empirical
autocorrelation for the data, and the solid lines represent the
theoretical autocorrelation constructed via MC simulation. The inset
presents the behavior of the tails in log-linear scale.}
\label{fig:auto}
\end{figure}


\begin{figure}
\centerline{\epsfig{file=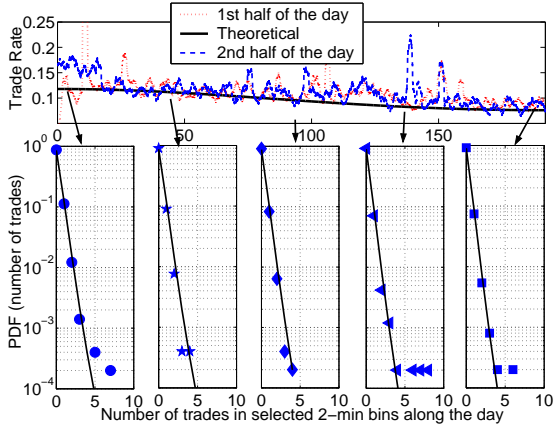,width=0.7\linewidth,angle=-90}}
\caption{\footnotesize\sf The top panel of the graph shows the daily
trading rate pattern [Eq.~(\ref{rateT})] for F. The bottom graphs
are the selected PDFs of 2-minute intervals along the trading day
for F; the symbols are data points, and the lines are the
theoretical PDF. Once the daily pattern is found (top panel), we fit
all 2-minute empirical PDFs along the day with one single parameter
$\sigma$ (lower panel).} \label{fig:pdf}
\end{figure}

After $r(0)$ and $A_s$ have been estimated, the remaining
fluctuation are assumed to be generated from the random process
$W(t)$ with intensity $\sigma$ [Eq.~(\ref{Vt})]. At this stage, we
only need to find $\tau_c$ (or $f_c$) [Eq.~(\ref{s2n})] to
demonstrate the presence of SR for a given stock.

We find $\tau_c$ indirectly through the autocorrelation function of
the number of trades $N(t)$. Notice that $\tau_c$ measures the
characteristic memory of the Gaussian noise $W(t)$ in
Equation~(\ref{Vt}). However, we cannot measure the autocorrelation
of $W(t)$ since $W(t)$ cannot be observed from the data directly.
Furthermore, we can achieve a 1-second resolution by using the
autocorrelation $N(t)$, since $N(t)$ is directly observed from the
market.

We find the autocorrelation of the number of trades for our model by
Monte Carlo (MC) simulating Equations (\ref{poisson}), (\ref{Rt}),
and (\ref{Vt}). We simulate a time series of the number of trades
$N(t)$ with a band-limited white noise $W(t)$ of corner frequency
$f_c$. For every $f_c$  (or $\tau_c$) there will be a time series of
the number of trades. For every such time series, we have an
autocorrelation function. The correlation time $\tau_c$ for a given
stock is the one which minimizes the mean square error between the
simulated autocorrelation and the empirical autocorrelation
function.

FIG. \ref{fig:auto} presents the autocorrelation of the number of
trades $N(t)$ for two stocks, ATI and F. We calculate the
autocorrelation function using all $21$ days of tick-by-tick data (a
total of $\approx 530000$ data points). These stocks represent the
typical autocorrelation for our $12$ stocks. We are able to model
autocorrelation function up to correlations of the order of $1\%$.
However, in agreement with the literature, some stocks clearly
present a slow decaying autocorrelation. Notice that F has a slow
decaying tail above the autocorrelation noise level ($\approx
3/\sqrt(530000) = 0.0041$). ATI, however, decays faster and has no
significant autocorrelation.

The presence of a persistent tail for F shows
that the autocorrelation could better modeled either by 2 exponentials
(two different $\tau_c$) or by a power law. The exponential autocorrelation function with a single $\tau_c$ is, therefore, a first order approximation.
However, such first order approximation will suffice to define SR. We note that is highly unlikely that $\tau_c$  will be small enough for a stock to
present a maximum in SNR and still have a large and significant autocorrelation tail. That is, most of the autocorrelation ($>99\%$) is generally
explained by a single exponential function. If a stock has a large autocorrelation tail, it will have a large $\tau_c$ as happens with F (and not with ATI).

\begin{figure}
\centerline{\epsfig{file=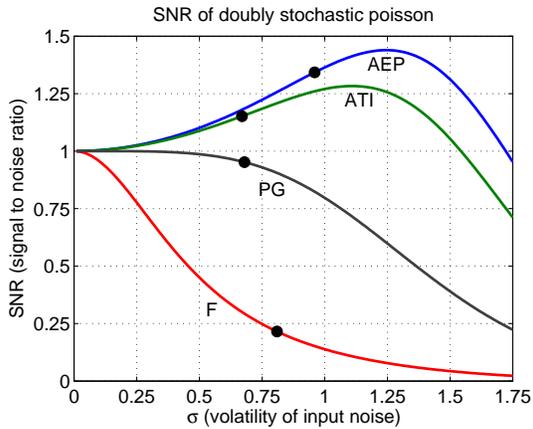,width=0.70\linewidth,angle=-90}}
\caption{\footnotesize\sf Output signal to noise ratio (SNR) of 4
trading stocks (AEP, ATI, PG, F) using Eq.~(\ref{s2n}). The circles
are the current value of $\sigma$. The figure shows that SR occurs
at different levels in AEP and AIT, but not in F and PG.}
\label{fig:snr}
\end{figure}

The noise intensity $\sigma$ defines the present trading environment
in the SNR graph [FIG.~(\ref{fig:snr})]. We find $\sigma$ by fitting
the theoretical probability density function (PDF) of the doubly
stochastic Poisson process to the empirical PDF \cite{Schonbucher2002}.
Note that $r(0)$ and $A_s$ were estimated previously via
Equation~(\ref{rateT}). We replace the previously estimated $r(0)$
and $A_s$ into Equation~(\ref{Rt}).
We continue with the approximation that every 2-minute interval is
described by doubly stochastic Poisson process with constant arrival
rate $r(0)\exp(A_s \cos(2\pi f_s t))$ perturbed by noise. Thus,
within each 2-minute interval, the noise can be described by the
log-normal distribution $\exp(W(t))$.
We apply numerical integration to calculate the integral of the
log-normal (which defines the random rate) and the Poisson (which
takes this random rate) to find the PDF of number of arrivals in
every 2-minute interval. Since the number of arrivals in 2-minute
intervals are small, we compensate the small number of arrivals by
fitting all the 2-minute intervals in a trading day at once to
obtain $\sigma$ by the mean square method.

FIG. \ref{fig:pdf} shows the daily trading pattern by binning a
trading day into $2$-minute intervals for F (other stocks present
similar results). The top part of the figure is the daily trading
pattern. For a cosine function, the first half of the day is the
mirror image of the second half. We put both halves on top of each
other. The solid thick line is obtained from our model. After
finding the parameterized mean, all the data points are fitted at
once to estimate the noise volatility $\sigma$. The bottom part of
the figure shows selected PDFs for few 2-min intervals along the
trading day. For instance, the first PDF represents the first and
the last 2-minute intervals of the trading day. FIG. \ref{fig:pdf}
shows that our simple model with a single $\sigma$ can fit the
empirical PDF up to $4$ orders of magnitude. This remarkable
agreement is found for all stocks studied along the entire trading
day.

Once $\sigma$ is obtained by our scheme, we have described the
trading environment using our model; thus, stocks that present SR can
be identified. FIG. \ref{fig:snr} shows the SNR as a function of
noise intensity. We have normalized all SNR curves such that the SNR
for $\sigma = 0$ is equal to $1$ for all stocks. Table \ref{table}
has the recovered values for the parameters of our model. We can see
that, in fact, stocks that satisfy $r(0)\tau_c < 0.25$ such as AEP
and ATI present a maximum on the SNR (and, hence, show SR). From our
sample of $12$ stocks, $1/2$ show SR. The largest $r(0)$ for the
stocks that present SR is $\approx 4$ trades per minute, whereas,
theoretically, we need to have $r(0)<15$ per minute (assuming that
$\tau_c =1$ sec is the lowest possible value). These results
indicate that the less-traded stocks are more likely to present SR.
Usually the less-traded stocks are stocks with smaller market
capitalization. Therefore, SR should be more important for small cap
stocks.

\begin{table}[bpt]
\begin{center}
\footnotesize
\begin{tabular}{p{.3in}cccccc}
\hline\hline & $r(0) (1/sec)$ & $A_s$ & $\sigma$ & $\tau_c (sec)$ &
$\tau_c r(0)$\\ \cline{2-6}
AEP & 0.0283 & 0.2880 & 0.9554 & 1.7000 &0.05&\\
ATI & 0.0146 & 0.1253 & 0.6688 & 4.7000 &0.05&\\
PG & 0.0946 & 0.2856 & 0.6839 & 2.6000 &0.26&\\
F & 0.0947 & 0.2194 & 0.8118 &  26.5000 &2.60&\\ \hline\hline
\end{tabular}%
\end{center}
\caption{\footnotesize\sf Parameter values for FIG. 4.}
\label{table}
\end{table}

The values in Table \ref{table} together with FIG. \ref{fig:snr}
describe the statistical trading environment. From the point of view
of a market observer, the signal of interest ($A_{S} \cos(2\pi f_c
 t)$) is being transmitted more efficiently (ATI and AEP) or less
efficiently (F and PG) with the help of noise present in the trading
environment. From a point of view of a market participant, both AEP and ATI are not at the SNR maximum. Therefore, adding noise by adding trades with a random arrival rate should increase the detection of the arrival rate "U"-shaped pattern. In the case of AEP, for instance, a $25\%$ increase in the noise intensity is beneficial to the periodic pattern.

\section{Conclusion}

We have proposed to model the number of stock trades inside of the trading day with a doubly stochastic Poisson process. With a doubly stochastic Poisson process, we are able to model both the deterministic periodic trend ("U"-shaped pattern) in the trade arrival rate as well as deviations from it due to noise. We show that such model describes well the empirical data and that non-dynamical stochastic resonance can be a relevant phenomena for the "U"-shaped pattern detection. SR as presented in this paper is defined as a weak regular oscillation that may be amplified by adding noise to the input signal. That means that the increase of noise trading (represented here with an increase in the random trade rate) can, in fact, augment the relevance of the exiting "U"-shaped pattern.

We envision that extensions of the model presented in this work could be possible for different economical time series. In our theoretical framework, the presence of a periodic pattern is a necessary condition for SR to be relevant. We know that such periodic patterns are, in fact, present in economics in general (see \cite{Franses1996,Heston2008,Garrido2005,Tsiakas2006} for additional examples), therefore making SR a potentially important phenomena.

\vspace*{.3in} \noindent\textbf{Acknowledgments.} We are grateful to
Peter Carr and Lee Deville for enlightening discussions. We thank
Samir Garzon for suggestions to improve our work as well as detailed
discussions. We thank Sergey Bezrukov for pointing us to useful
references, and Robert H. Smith School of Business at UMCP for
providing tick-by-tick data. We thank also Professor Jonathan
Goodman of the Courant Institute of Mathematical Sciences at New
York University for his encouragement and support.

\end{document}